\documentclass[preprint,twocolumn,5p]{elsarticle}

\usepackage{units}
\usepackage{xspace}
\usepackage{graphicx}
\usepackage{mathrsfs}
\usepackage{subfigure}
\usepackage{textpos}
\usepackage[breaklinks]{hyperref}

\usepackage{preprintcover} 
\PreprintCoverPaperTitle{Search for Events with Large Missing
  Transverse Momentum, Jets, and at Least Two Tau Leptons in 7~TeV
  Proton-Proton Collision Data with the ATLAS Detector} 
\PreprintIdNumber{CERN-PH-EP-2012-054} 
\PreprintCoverAbstract{A search for events with large missing
  transverse momentum, jets, and at least two tau leptons has been
  performed using 2~fb$^{-1}$ of proton-proton collision data at
  $\sqrt{s} = \unit[7]{TeV}$ recorded with the ATLAS detector at the
  Large Hadron Collider. No excess above the Standard Model background
  expectation is observed and a 95{\%}~CL upper limit on the visible
  cross section for new phenomena is set, where the visible cross
  section is defined by the product of cross section, branching
  fraction, detector acceptance and event selection efficiency. A
  95{\%}~CL lower limit of 32~TeV is set on the gauge-mediated
  supersymmetry breaking scale $\Lambda$ independent of $\tan\beta$.
  These limits provide the most stringent tests to date in a large
  part of the considered parameter space.}  
\PreprintJournalName{Physics Letters B} 

\biboptions{compress}

\newcommand{\Alpgen}{{\tt ALPGEN}\xspace}
\newcommand{\Jimmy}{{\tt JIMMY}\xspace}
\newcommand{\Mcatnlo}{{\tt MC@NLO}\xspace}
\newcommand{\Herwig}{{\tt HERWIG}\xspace}
\newcommand{\Herwigpp}{{\tt HERWIG++}\xspace}
\newcommand{\Isajet}{{\tt ISAJET}\xspace}
\newcommand{\Prospino}{{\tt PROSPINO}\xspace}
\newcommand{\Tauola}{{\tt TAUOLA}\xspace}
\newcommand{\Photos}{{\tt PHOTOS}\xspace}
\newcommand{\Pythia}{{\tt PYTHIA}\xspace}

\newcommand{\antibar}[1]{\ensuremath{#1\bar{#1}}}
\newcommand{\ttbar}{\antibar{t}\xspace}
\newcommand{\neutralino}{\ensuremath{\tilde{\chi}^{0}_{1}}\xspace}
\newcommand{\gravitino}{\ensuremath{\tilde{G}}\xspace}

\newcommand{\stau}{\ensuremath{\tilde{\tau}_1}\xspace}
\newcommand{\slepton}{\ensuremath{\tilde{\ell}_R}\xspace}
\newcommand{\sneutrino}{\ensuremath{\tilde{\nu}}\xspace}

\newcommand{\Nfive}{\ensuremath{N_{5}}\xspace}
\newcommand{\Mmess}{\ensuremath{M_{\mathrm{mess}}}\xspace}
\newcommand{\Cgrav}{\ensuremath{C_{\mathrm{grav}}}\xspace}

\newcommand{\CL}{CL\xspace}
\newcommand{\met}{\ensuremath{E_{\mathrm{T}}^{\mathrm{miss}}}\xspace}
\newcommand{\ptm}{\ensuremath{p_{\mathrm{T}}^{\mathrm{miss}}}\xspace}
\newcommand{\meff}{\ensuremath{m_{\mathrm{eff}}}\xspace}

\newcommand{\pt}{\ensuremath{p_{\mathrm{T}}}\xspace}

\newcommand{\mass}[1]{\ensuremath{m_{#1}}}
\newcommand{\mT}{\ensuremath{\mass{\mathrm T}}\xspace}
\newcommand{\mTT}{\ensuremath{\mass{\mathrm T}^{\tau_1}+\mass{\mathrm T}^{\tau_2}}\xspace}

\newcommand{\Section}{Section}
\newcommand{\Fig}{Fig.\xspace}
\newcommand{\Ref}{Ref.\xspace}
\newcommand{\Refs}{Refs.\xspace}

\newcommand{\ipb}{\mbox{pb$^{-1}$}}
\newcommand{\ifb}{\mbox{fb$^{-1}$}}

\newcommand{\integLumi}{\unit[2]{\ifb}\xspace}
\newcommand{\integLumiE}{\unit[$(2.05\pm 0.08)$]{\ifb}\xspace}
\newcommand{\nData}{3\xspace}
\newcommand{\nBackground}{\ensuremath{5.3\pm1.3(\mathrm{stat})\pm2.2(\mathrm{sys})}\xspace}
\newcommand{\nGMSBBenchmark}{\ensuremath{20.8\pm3.4(\mathrm{stat})\pm3.6(\mathrm{sys})\pm3.3(\mathrm{theo})}\xspace}
\newcommand{\NCLs}{\ensuremath{5.9}\xspace}
\newcommand{\NCLsExp}{\ensuremath{7.0}\xspace}
\newcommand{\vXSLimit}{\unit[2.9]{fb}\xspace}
\newcommand{\vXSLimitExp}{\unit[3.4]{fb}\xspace}
\newcommand{\GMSBLimitBestCLs}{\unit[47]{TeV}\xspace}
\newcommand{\GMSBLimitBestTanBetaCLs}{37\xspace}
\newcommand{\GMSBLimitCLs}{\unit[32]{TeV}\xspace}

\journal{Physics Letters B}

\begin{document}

\begin{frontmatter}

  \title{Search for Events with Large Missing Transverse Momentum,
    Jets, and at Least Two Tau Leptons in \unit[7]{TeV} Proton-Proton
    Collision Data with the ATLAS Detector}

\author{The ATLAS Collaboration}

\begin{abstract}
  A search for events with
  large missing transverse momentum, jets, and at least two tau
  leptons has been performed using \integLumi of proton-proton
  collision data at $\sqrt{s} = \unit[7]{TeV}$ recorded with the ATLAS
  detector at the Large Hadron Collider. No excess above the Standard
  Model background expectation is observed and a 95{\%}~CL upper limit
  on the visible cross section for new phenomena is set, where the
  visible cross section is defined by the product of cross section,
  branching fraction, detector acceptance and event selection
  efficiency. A \unit[95]{\%}~CL lower limit of \GMSBLimitCLs is set
  on the gauge-mediated supersymmetry breaking (GMSB) scale $\Lambda$
  independent of $\tan\beta$. These limits provide the most stringent
  tests to date in a large part of the considered parameter space.
\end{abstract}

\end{frontmatter}

\section{Introduction}

Supersymmetry~(SUSY)~\cite{Golfand:1971iw,Neveu:1971rx,Ramond:1971gb,Volkov:1973ix,Wess:1974tw}
introduces a symmetry between fermions and bosons, resulting in a SUSY
partner (sparticle) for each Standard Model (SM) particle with
identical mass and quantum numbers except a difference by half a unit
of spin.  As none of these sparticles have been observed, SUSY must be
a broken symmetry if realised in nature.  Assuming $R$-parity
conservation~\cite{Fayet:1977yc,Farrar:1978xj}, sparticles are
produced in pairs.  These would then decay through cascades involving
other sparticles until the lightest SUSY particle (LSP) is produced,
which is stable.

Minimal gauge-mediated supersymmetry breaking
(GMSB)~\cite{Dine:1981gu, AlvarezGaume:1981wy, Nappi:1982hm,
  Dine:1993yw, Dine:1994vc, Dine:1995ag} models can be described by
six parameters: the SUSY breaking mass scale felt by the low-energy
sector ($\Lambda$), the messenger mass (\Mmess), the number of SU(5)
messengers (\Nfive), the ratio of the vacuum expectation values of the
two Higgs doublets ($\tan\beta$), the Higgs sector mixing parameter
($\mu$) and the scale factor for the gravitino mass (\Cgrav). In this
analysis $\Lambda$ and $\tan\beta$ are treated as free parameters and
the other parameters are fixed to $\Mmess=\unit[250]{TeV}$,
$\Nfive=3$, $\mu>0$ and $\Cgrav=1$, similar to other GMSB benchmark
points in the literature, e.g. G2a~\cite{Hinchliffe:1998ys} and
SPS7~\cite{Allanach:2002nj}. The $\Cgrav$ parameter determines the
lifetime of the next-to-lightest SUSY particle~(NLSP). For $\Cgrav=1$
the NLSP decays promptly ($c\tau_{\mathrm{NLSP}} < \unit[0.1]{mm}$).
With these parameters, the production of squark and/or gluino pairs is
expected to dominate at the present Large Hadron Collider (LHC)
energy. These sparticles decay directly or through cascades into
the NLSP, which subsequently decays to the LSP.
In GMSB models, the LSP is the very light gravitino (\gravitino). Due
to the gravitino's very small mass of $\mathcal{O}(\unit{keV})$, the
NLSP is the only sparticle decaying into the LSP. This leads to
multiple jets and missing transverse momentum (\met) in the final
states.
The experimental signature is then largely determined by the nature of
the NLSP, which can be either the lightest stau (\stau), a right
handed slepton (\slepton), the lightest neutralino (\neutralino), or
a sneutrino (\sneutrino), leading to final states containing taus,
light leptons ($\ell=e,\mu$), photons, b-jets, or neutrinos.
For $\Nfive=3$ the \stau and \slepton NLSPs become more dominant
compared to lower values of \Nfive. At large values of $\tan\beta$,
the \stau is the NLSP for most of the parameter space, which leads to
final states containing between two and four tau leptons.
In the so-called CoNLSP~\cite{Ambrosanio:2000ik} region, the mass
difference between the \stau and the \slepton is smaller than the tau
lepton mass such that both sparticles decay directly into the LSP and
are therefore NLSP.

This Letter reports on the search for events with large \met, jets,
and at least two hadronically decaying tau leptons.  The analysis has
been performed using \integLumi of proton-proton ($pp$) collision data
at $\sqrt{s}=\unit[7]{TeV}$ recorded with the ATLAS detector at the
LHC between March and August 2011. Although the analysis is sensitive
to a wide variety of models for physics beyond the Standard Model, the
results shown here are interpreted in the context of a minimal GMSB
model.
The three LEP Collaborations ALEPH~\cite{Heister:2002vh},
DELPHI~\cite{Abdallah:2002rd} and OPAL~\cite{Abbiendi:2005gc} studied
\stau pair production, with the subsequent decay
$\stau\to\tau\gravitino$ in the minimal GMSB model. The best limits
are set by the OPAL Collaboration and \stau \mbox{NLSPs} with masses
below \unit[87.4]{GeV} are excluded. A limit on the SUSY breaking
mass scale $\Lambda$ of \unit[26]{TeV} was set for $\Nfive=3$,
$\Mmess=\unit[250]{TeV}$, independent of $\tan\beta$ and the NLSP lifetime.
The CMS Collaboration searched for new physics in same-sign ditau
events~\cite{Chatrchyan:2011wba} and multi-lepton events including
ditaus~\cite{Chatrchyan:2011ff} using \unit[35]{\ipb} of data, but the
minimal GMSB model was not considered.
A search for supersymmetry in final states containing at
least one hadronically decaying tau lepton, missing transverse 
momentum and jets with the ATLAS detector is presented in another 
Letter~\cite{onetaupaper}.

\section{ATLAS detector}

The ATLAS detector~\cite{Aad:2008zzm} is a multi-purpose apparatus
with a forward-backward symmetric cylindrical geometry and nearly
4$\pi$ solid angle coverage. The inner tracking detector~(ID) consists
of a silicon pixel detector, a silicon strip detector and a
transition radiation tracker. The ID is surrounded by a thin
superconducting solenoid providing a \unit[2]{T} magnetic field and by
fine-granularity lead/liquid-argon (LAr) electromagnetic calorimeters.
An iron/scintillating-tile calorimeter provides hadronic coverage in
the central rapidity\footnote{ATLAS uses a right-handed coordinate
  system with its origin at the nominal interaction point (IP) in the
  centre of the detector and the $z$-axis along the beam pipe. The
  $x$-axis points from the IP to the centre of the LHC ring and the
  $y$-axis points upward. Cylindrical coordinates $(R,\phi)$ are used
  in the transverse plane, $\phi$ being the azimuthal angle around the
  beam pipe. The pseudorapidity is defined in terms of the polar angle
  $\theta$ as $\eta=-\ln\tan(\theta/2)$.}  range. The endcap and
forward regions are instrumented with liquid-argon calorimeters for
both electromagnetic and hadronic measurements.
An extensive muon spectrometer system that incorporates large
superconducting toroidal magnets surrounds the calorimeters.

\section{Simulated samples}

Monte Carlo~(MC) simulations are used to extrapolate backgrounds
from control regions (CRs) to the signal region (SR) and to evaluate
the selection efficiencies for the SUSY models considered. Samples of
$W$ and $Z/\gamma^*$ production with accompanying jets are simulated
with \Alpgen~\cite{Mangano:2002ea}, using {\tt
  CTEQ6L1}~\cite{Pumplin:2002vw} parton density functions (PDFs). Top
quark pair production, single top production and diboson pair
production are simulated with
\Mcatnlo~\cite{Frixione:2002ik,Frixione:2003ei,Frixione:2005vw} and
the next-to-leading order (NLO) PDF set {\tt
  CTEQ6.6}~\cite{Nadolsky:2008zw}. 
Fragmentation and hadronisation are performed with
\Herwig~\cite{Corcella:2000bw}, using \Jimmy~\cite{Butterworth:1996zw}
for the underlying event simulation and the ATLAS {\tt MC10} parameter
tune~\cite{mc10}. \Tauola~\cite{tauola1,tauola2} and
\Photos~\cite{photos} are used to model the decays of $\tau$ leptons
and the radiation of photons, respectively.
The production of multi-jet events is simulated with \Pythia
6.4.25~\cite{Sjostrand:2006aa} using the AMBT1 tune~\cite{mc10AMBT1}
and {\tt MRST2007 LO${}^{*}$}~\cite{Sherstnev:2007nd} PDFs.
For the minimal GMSB model considered in this analysis, the SUSY mass
spectra are calculated using \Isajet~7.80~\cite{isajet}. The MC signal
samples are produced using \Herwigpp~2.4.2~\cite{herwig++} with {\tt
  MRST2007 LO${}^{*}$} PDFs. NLO cross sections are calculated using
\Prospino~2.1~\cite{prospino2,Beenakker:1996ch,Beenakker:1997ut,Beenakker:1999xh,Spira:2002rd,Plehn:2004rp}.
All samples are processed through the {\tt GEANT4}-based
simulation~\cite{geant4} of the ATLAS detector~\cite{Aad:2010wq}.  The
variation of the number of $pp$ interactions per bunch crossing
(pile-up) as a function of the instantaneous luminosity is taken into
account by modeling the simulated number of overlaid minimum bias
events according to the observed distribution of the number of pile-up
interactions in data, with an average of $\sim6$ interactions.

\section{Object reconstruction}

Jets are reconstructed using the anti-$k_t$ jet clustering
algorithm~\cite{Cacciari:2008gp} with radius parameter $R=0.4$. Their
energies are calibrated to correct for
calorimeter non-compensation, upstream material and other
effects~\cite{Aad:2011he}.  Jets are required to have transverse
momentum (\pt) above \unit[20]{GeV} and $|\eta|<2.5$.

Muons are identified as tracks in the ID matched to track segments in
the stand-alone muon spectrometer, while electrons are identified as
isolated tracks with a corresponding energy deposit in the
electromagnetic calorimeter. The selection criteria applied to muons
and ``medium'' quality electrons are described in more detail in Refs.
\Refs~\cite{Aad:2011cw} and \cite{Aad:2011mk}, respectively.

The measurement of the missing transverse momentum two-dimensional
vector \ptm (and its magnitude \met) is based on the transverse
momenta of identified jets, electrons, muons and all calorimeter
clusters with $|\eta| < 4.5$ not associated to such
objects~\cite{Aad:2011re}. For the purpose of the measurement of \met, 
taus are not distinguished from jets.

In this search, only hadronically decaying taus are considered.
The tau reconstruction is seeded from anti-$k_t$ jets with 
$\pt > \unit[10]{GeV}$. An $\eta$- and $\pt$-dependent energy
calibration to the hadronic tau energy scale is applied. 
Hadronic tau identification is based on observables sensitive to the
transverse and longitudinal shape of the calorimeter shower and on
tracking information, combined in a boosted decision tree~(BDT)
discriminator~\cite{ATLAS-CONF-2011-152}.
Transition radiation and calorimeter information is used to veto
electrons misidentified as taus.  A tau candidate must have
$\pt>\unit[20]{GeV}$, $|\eta|<2.5$, and one or three associated tracks
of $\pt>\unit[1]{GeV}$ with a charge sum of $\pm1$. The efficiency of
the BDT tau identification (the ``loose'' working point in
\Ref~\cite{ATLAS-CONF-2011-152}), determined using $Z \rightarrow \tau
\tau$ events, is about \unit[60]{\%}, independent of \pt, with
a jet background rejection factor of $20-50$.

During a part of the data-taking period, an electronics failure in the
LAr barrel EM calorimeter created a dead region in the second and
third layers, corresponding to approximately
$1.4\times\unit[0.2]{radians}$ in $\Delta \eta \times \Delta \phi$.
Electron and tau candidates falling in this region are discarded.
A correction to the jet energy is made using the energy depositions in
the cells neighbouring the dead region; events
having at least one jet for which the energy after correction is above
\unit[30]{GeV} are discarded, resulting in a loss of $\sim6${\%} of
the data sample.

\begin{figure}[t]
  \begin{center}
    \includegraphics[width=0.48\textwidth]{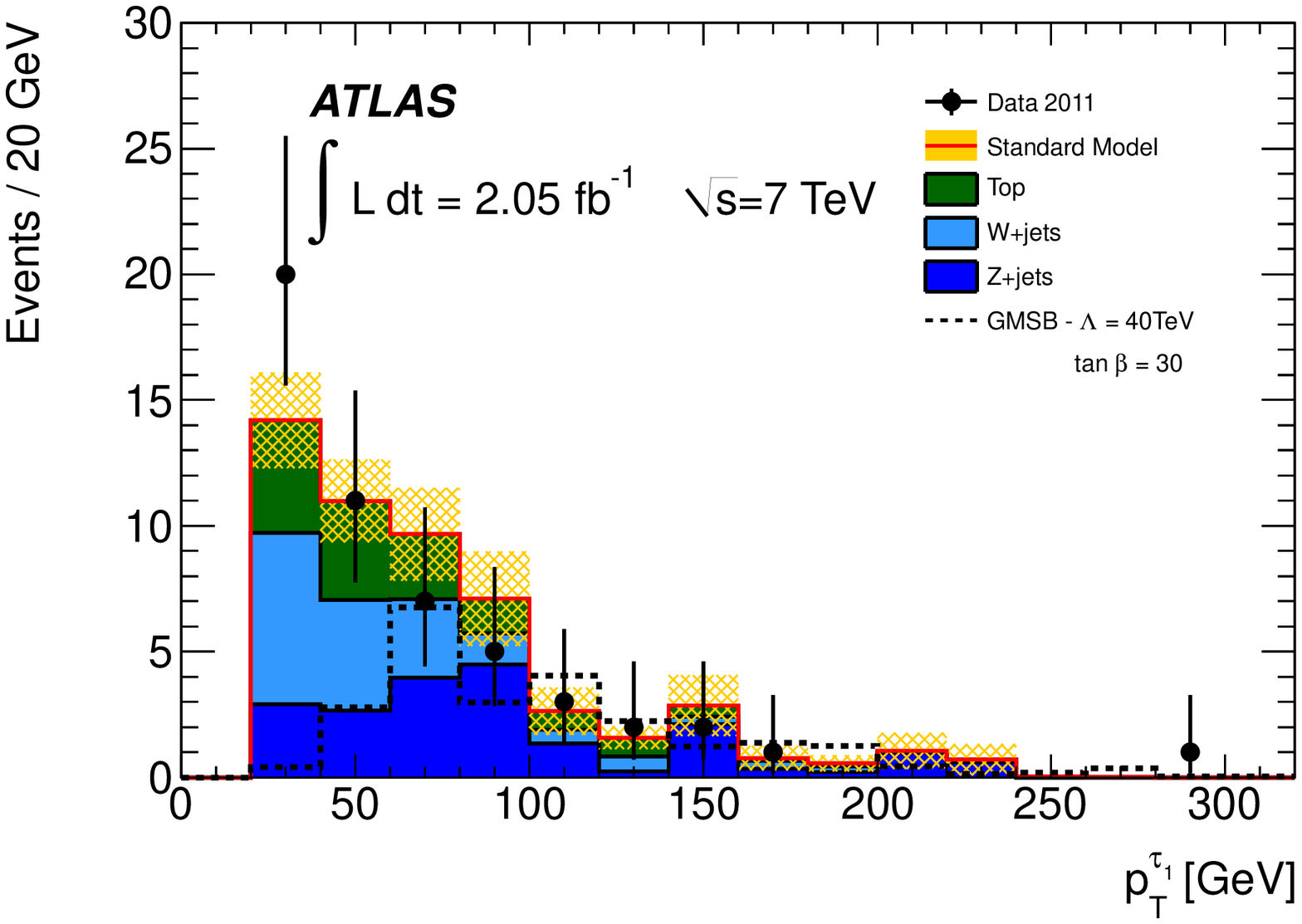}
  \end{center}
  \caption{The \pt spectrum of the leading tau candidates in
    data (points, statistical uncertainty only) and the estimated SM
    background after the pre-selection of candidate events, soft multi-jet
    rejection and the requirement of two or more taus and no light leptons. 
    The band centered around
    the total SM background indicates the statistical uncertainty.
    Also shown is the expected signal from a typical GMSB
    ($\Lambda=\unit[40]{TeV}, \tan\beta=30$) sample.
    \label{fig:taupt}
  }
\end{figure}

\section{Data analysis}
\label{dataAnalysis}

The analysed data sample, after applying beam, detector and
data-quality requirements, corresponds to an integrated luminosity of
\integLumiE~\cite{lumi2011,Aad:2011dr}. Candidate events are pre-selected
by a trigger requiring a leading jet, i.e. the jet having the highest
transverse momentum of all jets in the event, with
$\pt>\unit[75]{GeV}$, measured at the raw electromagnetic scale, and
$\met>\unit[45]{GeV}$~\cite{Aad:2011xs}.
In the offline analysis, these events are required to have a
reconstructed primary vertex with at least five tracks, a leading jet
with $\pt>\unit[130]{GeV}$ and $\met>\unit[130]{GeV}$. These
requirements ensure a uniform trigger efficiency that exceeds
\unit[98]{\%}.

Pre-selected events are then required to have at least two identified tau
candidates and must not contain any electron or muon candidates with
transverse momenta above \unit[20]{GeV} or \unit[10]{GeV},
respectively. To suppress soft multi-jet events, a second jet with
$\pt>\unit[30]{GeV}$ is required. The \pt spectrum of the leading tau
candidate after pre-selection of candidate events, soft multi-jet rejection and the
requirement of two or more taus and no light leptons is shown in \Fig~\ref{fig:taupt}.

This selection rejects almost all soft multi-jet background events.
Remaining multi-jet events, where highly energetic jets are
mis-measured, are rejected by requiring the azimuthal angle between
the missing transverse momentum and either of the two leading jets
$\Delta\phi(\ptm, \mathrm{jet}_{1,2})$ to be larger than
\unit[0.4]{radians}.

\begin{figure}[h!]
  \begin{center}
    \subfigure[~\meff distribution after the $\Delta\phi$ requirement]{\includegraphics[width=0.48\textwidth]{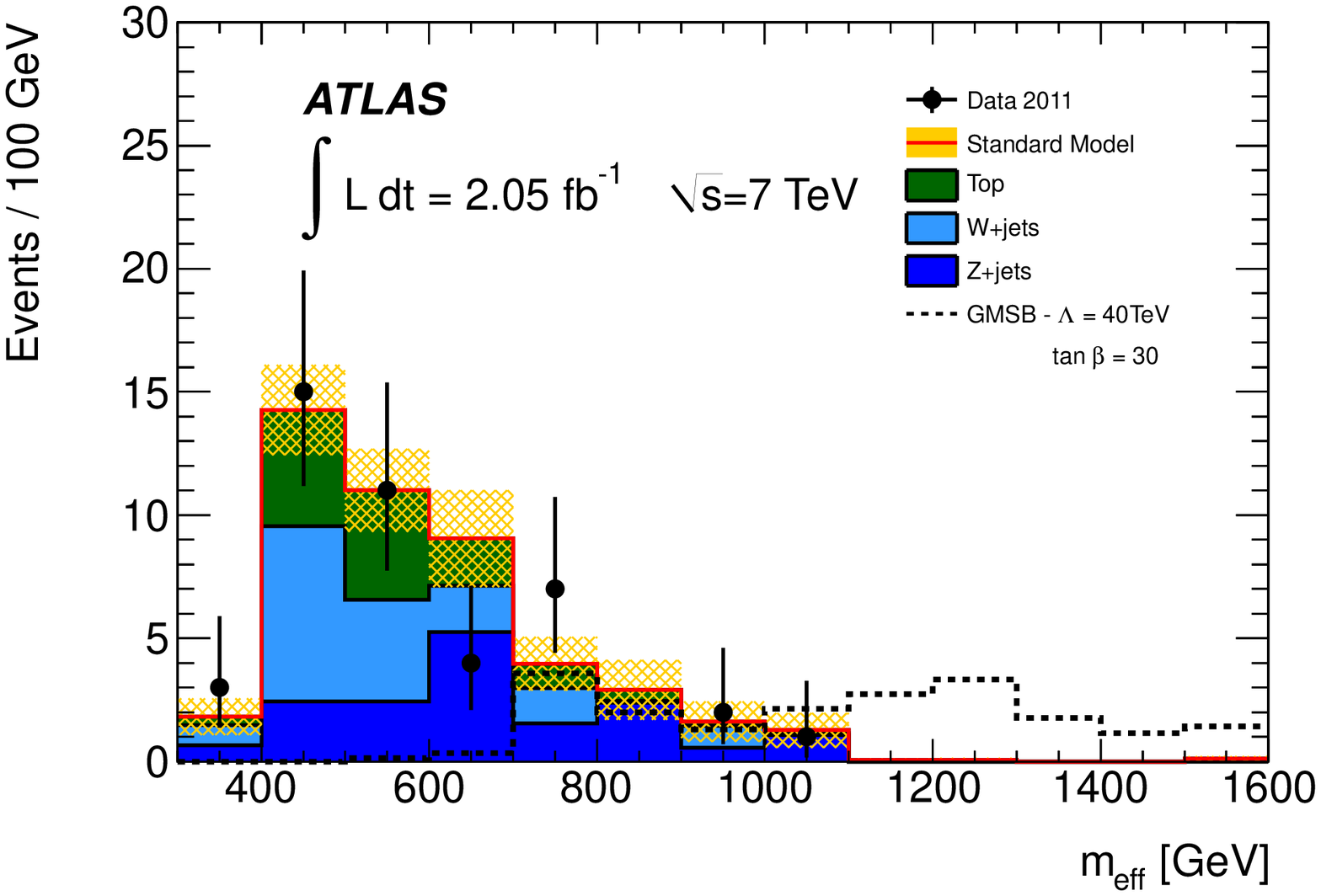}}
    \subfigure[~\mTT distribution after the \meff requirement]{\includegraphics[width=0.48\textwidth]{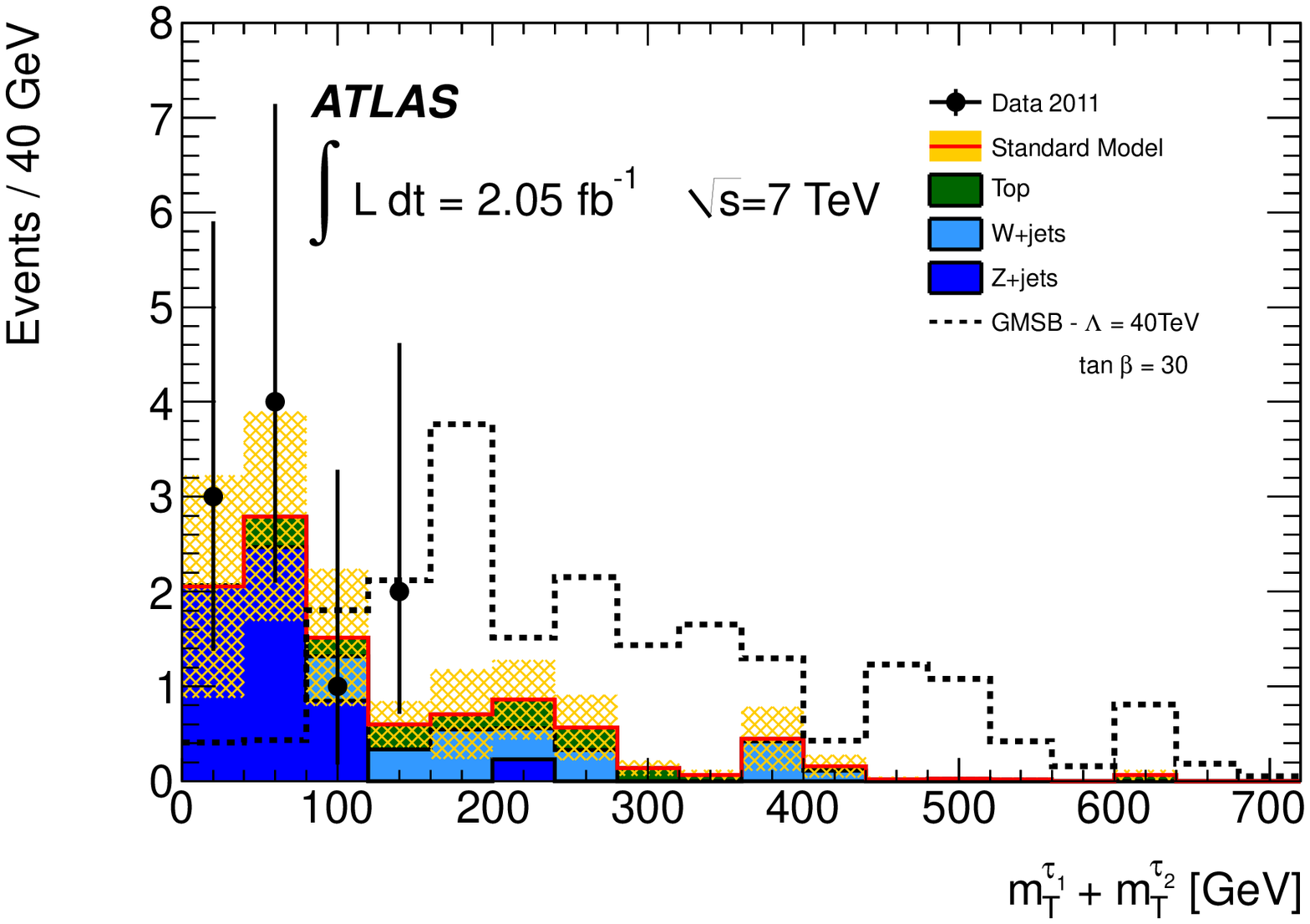}}
  \end{center}
  \caption{Distributions of variables used for the signal region
    definition in data (points, statistical uncertainty only) and the
    estimated SM background after the pre-selection of candidate events, soft multi-jet rejection and 
    the requirement of two or more taus and no light leptons.
    The yellow band centered around the total SM background
    indicates the statistical uncertainty.
    Also shown is the expected signal from a typical GMSB ($\Lambda=\unit[40]{TeV}, \tan\beta=30$) sample. 
  }
  \label{fig:2tauResults_meff}
\end{figure}

The SR is defined by requiring $\meff>\unit[700]{GeV}$ and
$\mTT>\unit[80]{GeV}$, where \meff is the effective mass\footnote{The
  effective mass \meff is calculated as the sum of \met and the
  magnitude of the transverse momenta of the two highest-\pt jets and
  all selected taus.} and \mTT is the sum of the transverse
masses\footnote{The transverse mass \mT formed by \met and the \pt of
  the tau lepton ($\tau$) is defined as
  $\mT=\sqrt{2\pt^{\tau}\met(1-\cos(\Delta\phi(\tau,\ptm)))}$.} of the
two leading tau candidates.
The \meff distribution after the $\Delta\phi(\ptm,
\mathrm{jet}_{1,2})$ requirement and the \mTT distribution after the
\meff requirement are shown in \Fig~\ref{fig:2tauResults_meff}.
After applying all the analysis requirements, \nData events are
selected in the data.

\section{Background estimation}

The dominant backgrounds in the SR arise from top-pair plus single top
events (here generically indicated as \ttbar), $W\to\tau\nu_{\tau}$ events 
and $Z \rightarrow \tau \tau$ events. While the latter comprises final states 
with two true taus, which are well described in the simulation, the $W$ and 
\ttbar background consist of events in which one real tau is correctly
reconstructed and the other tau candidates are mis-reconstructed from
hadronic activity in the final state.
Since mis-identified taus are not well described in the MC, the background 
contribution from \ttbar and $W\to\tau\nu_{\tau}$ is determined simultaneously 
in a CR defined by inverting the \meff cut. Owing to the requirement on $\Delta\phi$
and of two or more taus, this CR has negligible contamination from 
multi-jet events. Moreover, a totally negligible contribution is expected in
this CR from signal events. The MC overestimates the number of events in the CR
compared to data, due to mis-modeling of tau misidentification
probabilities. MC studies show that the tau misidentification
probability is, to a good approximation, independent of \meff, so that
the measured ratio of the data to MC event yields in the CR can be
used to correct the MC background prediction in the SR.

In a similar way, the multi-jet background expectation is computed
in a multi-jet dominated CR defined by inverting the $\Delta\phi$ and
\meff cuts. In addition, $\met/\meff<0.4$ is required to increase the
purity of this CR sample. The extrapolated contribution of this
background source to the SR is found to be negligible.

\section{Systematic uncertainties on the background}
\label{sec:syst}

The theoretical uncertainty on the MC-based corrected extrapolation of
the $W$ and \ttbar backgrounds from the CR into the SR is estimated
using alternative MC samples obtained by varying the renormalisation
and factorisation scales, the functional form of the factorisation
scale and the matching threshold in the parton shower process.
An uncertainty of \unit[14]{\%} is estimated from this procedure.
Moreover, an uncertainty of \unit[23]{\%} is associated to the
normalisation factor derived in the CR. This uncertainty is estimated
by repeating the normalisation to data independently for $W$ and
\ttbar.
Systematic uncertainties on the jet energy scale and jet energy
resolution~\cite{Aad:2011he} are applied in MC to the selected jets
and propagated throughout the analysis, including to \met. The
difference in the number of expected background events obtained with
the nominal MC simulation after applying these changes is taken as
the systematic uncertainty and corresponds to \unit[18]{\%} each.
The effect of the tau energy scale uncertainly on the
expected background is estimated in a similar way and amounts to
\unit[7]{\%}. The uncertainties from the jet and tau energy
scale are treated as fully correlated.
The tau identification efficiency uncertainties on the background
depends on the tau identification algorithm, the kinematics of the
$\tau$ sample and the number of associated tracks. The systematic
uncertainties associated to the tau identification and
misidentification are found to be \unit[2.5]{\%} and \unit[0.5]{\%},
respectively. For the \ttbar and $W$ backgrounds, these uncertainties
are absorbed into the normalisation.
The systematic uncertainty associated to pile-up simulation in MC is
\unit[1]{\%}.
The normalization of the Z+jets and diboson backgrounds is affected by 
the uncertainty of \unit[3.7]{\%} on the luminosity measurement~\cite{lumi2011,Aad:2011dr}.
This results in a \unit[0.8]{\%} uncertainty on the total background.
The contributions from the different systematic uncertainties result
in a total background systematic uncertainty of \unit[41]{\%}.

In total \nBackground background events are expected where the first
uncertainty is statistical and includes the statistical component of
the background correction factor uncertainty and the second is
systematic.
Roughly half of the background is composed of \ttbar events and the
other half is evenly split into $W$ and $Z$ events with accompanying
jets.
%

\section{Signal efficiencies and systematic uncertainties}

GMSB signal samples were generated on a grid ranging from $\Lambda =
\unit[10]{TeV}$ to $\Lambda = \unit[80]{TeV}$ and from $\tan\beta = 2$
to $\tan\beta = 50$. The number of selected events decreases
significantly with increasing $\Lambda$ due to the reduced cross
section.  The cross section drops from $\unit[100]{pb}$
for $\Lambda = \unit[15]{TeV}$ to $\unit[5.0]{fb}$ for $\Lambda =
\unit[80]{TeV}$. The selection efficiency is highest ($\approx3\%$)
for high $\tan\beta$ and lower $\Lambda$ values, including in the region
of the GMSB4030 point ($\Lambda = 40$, $\tan\beta=30$) which is near
the expected limit. It drops to \unit[0.2]{\%} in the non-\stau NLSP
regions and for high $\Lambda$ values. This is primarily a consequence
of the light lepton veto and the requirement of two hadronically
decaying taus, respectively.

The total systematic uncertainty on the signal selection from the
systematic uncertainties discussed in \Section~\ref{sec:syst} ranges
between \unit[7.5]{\%} and \unit[36]{\%} over the GMSB grid. The
statistical uncertainty from the limited size of the MC signal samples
is of the order of \unit[20]{\%}, with variations between
\unit[7.6]{\%} and \unit[59]{\%} at the edges of the accessible signal
range. 
Theory uncertainties related to the GMSB cross section predictions are
estimated through variations of the factorisation and renormalisation
scales in the NLO \Prospino calculation between half and twice their
default values, by considering variations in $\alpha_s$, and by
considering PDF uncertainties using the {\tt CTEQ6.6M} PDF error
sets~\cite{Stump:2003yu}.
These uncertainties are calculated for individual SUSY production
processes and for each model point, leading to overall theoretical
cross section uncertainties between \unit[6.5]{\%} and \unit[22]{\%}.
Altogether this yields \nGMSBBenchmark signal events for the GMSB4030
point.

\section{Results}

\begin{figure}[t]
  \begin{center}
    \includegraphics[width=0.48\textwidth]{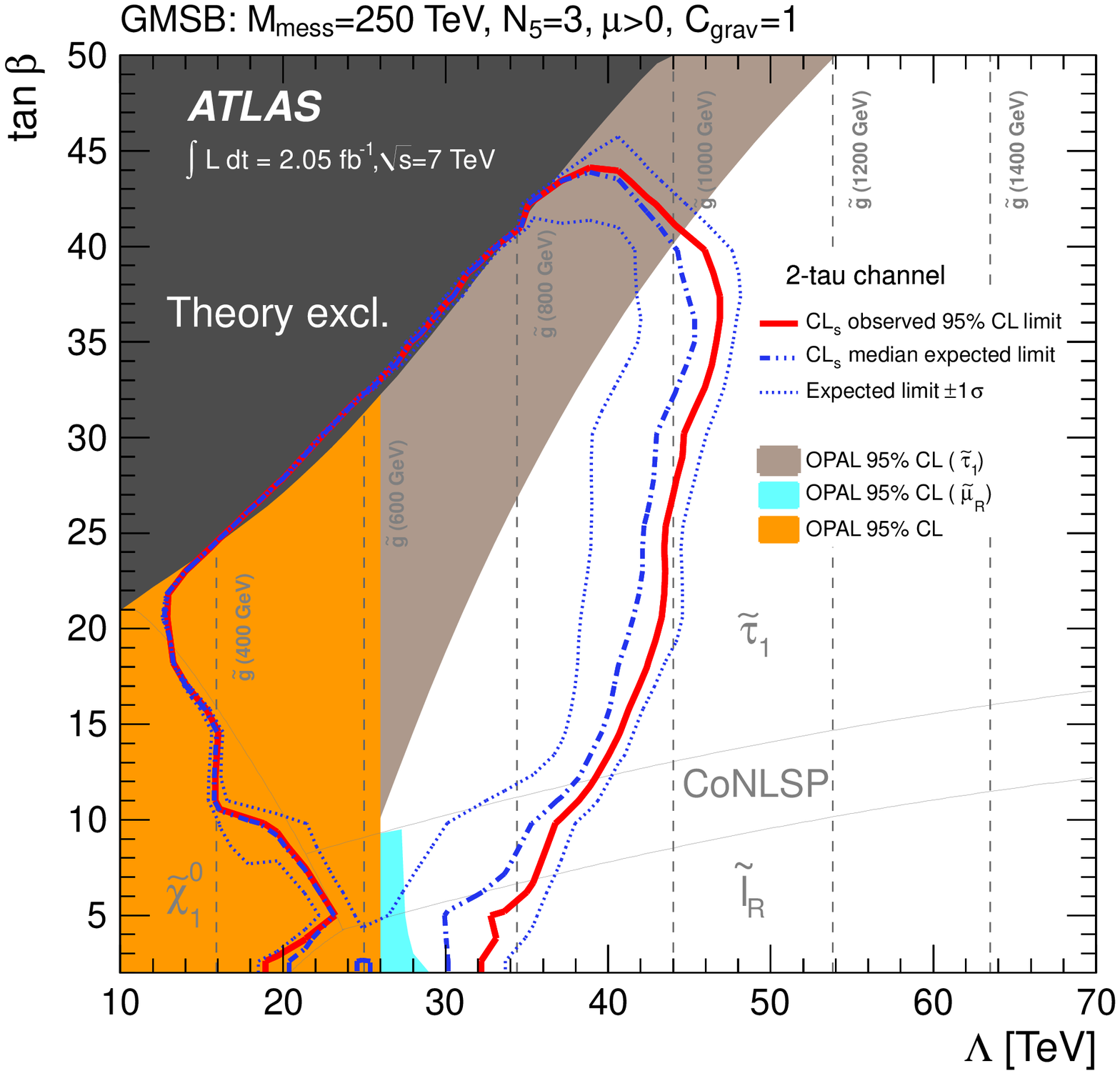}
  \end{center}
  \caption{Expected and observed \unit[95]{\%}~\CL limits on the
    minimal GMSB model parameters $\Lambda$ and $\tan\beta$. 
    The dark grey area indicates the region which
    is theoretically excluded due to unphysical sparticle mass values.
    The different NLSP regions are indicate. In the CoNLSP region the \stau and the \slepton are the NLSP.
    Additional model parameters are $\Mmess = \unit[250]{TeV}$,
    $\Nfive=3$, $\mu>0$ and $\Cgrav=1$. The previous
    OPAL~\protect\cite{Abbiendi:2005gc} limits are also shown.  }
  \label{fig:gmsb:limit}
\end{figure}

Based on the observation of \nData events in the SR
and a background expectation of \nBackground events, an upper limit
of \NCLs~(\NCLsExp) events observed (expected) is set at
\unit[95]{\%}~Confidence Level~(\CL)
on the number of events from any scenario
of physics beyond the SM, using the profile likelihood and
$CL_s$ method~\cite{Read:2002hq}. Uncertainties on the background and
signal expectations are treated as Gaussian-distributed nuisance
parameters in the likelihood fit.
This limit translates into a \unit[95]{\%} observed (expected) upper
limit of \vXSLimit~(\vXSLimitExp) on the visible cross section for new
phenomena, defined by the product of cross section, branching
fraction, acceptance and efficiency for the selections defined in Section~\ref{dataAnalysis}.
The resulting expected and observed \unit[95]{\%}~\CL limits on the
GMSB model parameters $\Lambda$ and $\tan\beta$ are shown in
\Fig~\ref{fig:gmsb:limit}, including the lower limits from
OPAL~\cite{Abbiendi:2005gc} for comparison.  These limits are
calculated including all experimental and theoretical uncertainties on
the background and signal expectations. Excluding the theoretical
uncertainties on the signal cross section from the limit calculation
has a negligible effect on the limits obtained.
The best exclusion is set for $\Lambda = \GMSBLimitBestCLs$ and
$\tan\beta=\GMSBLimitBestTanBetaCLs$. 
The results extend previous limits and values of $\Lambda <
\GMSBLimitCLs$ are now excluded at \unit[95]{\%}~\CL, independent of
$\tan\beta$.
%

\section{Conclusions}

A search for events with two or more hadronically decaying tau
leptons, large \met and jets is performed using \integLumi of
$\sqrt{s} = \unit[7]{TeV}$ $pp$ collision data recorded with the ATLAS
detector at the LHC.  Three events are found, consistent with the
expected SM background.
The results are used to set a model-independent \unit[95]{\%}~\CL
upper limit of \NCLs events from new phenomena, corresponding to an
upper limit on the visible cross section of \vXSLimit.
Limits on the model parameters are set for a minimal GMSB model.
The limit on the SUSY breaking scale $\Lambda$ of \GMSBLimitCLs is
determined, independent of $\tan\beta$. It increases up to
\GMSBLimitBestCLs for $\tan\beta=\GMSBLimitBestTanBetaCLs$.
These results provide the most stringent tests in a large part of the
parameter space considered to date, improving the previous best limits.

\section*{Acknowledgements}




We thank CERN for the very successful operation of the LHC, as well as the
support staff from our institutions without whom ATLAS could not be
operated efficiently.

We acknowledge the support of ANPCyT, Argentina; YerPhI, Armenia; ARC,
Australia; BMWF, Austria; ANAS, Azerbaijan; SSTC, Belarus; CNPq and FAPESP,
Brazil; NSERC, NRC and CFI, Canada; CERN; CONICYT, Chile; CAS, MOST and NSFC,
China; COLCIENCIAS, Colombia; MSMT CR, MPO CR and VSC CR, Czech Republic;
DNRF, DNSRC and Lundbeck Foundation, Denmark; EPLANET and ERC, European Union;
IN2P3-CNRS, CEA-DSM/IRFU, France; GNAS, Georgia; BMBF, DFG, HGF, MPG and AvH
Foundation, Germany; GSRT, Greece; ISF, MINERVA, GIF, DIP and Benoziyo Center,
Israel; INFN, Italy; MEXT and JSPS, Japan; CNRST, Morocco; FOM and NWO,
Netherlands; RCN, Norway; MNiSW, Poland; GRICES and FCT, Portugal; MERYS
(MECTS), Romania; MES of Russia and ROSATOM, Russian Federation; JINR; MSTD,
Serbia; MSSR, Slovakia; ARRS and MVZT, Slovenia; DST/NRF, South Africa;
MICINN, Spain; SRC and Wallenberg Foundation, Sweden; SER, SNSF and Cantons of
Bern and Geneva, Switzerland; NSC, Taiwan; TAEK, Turkey; STFC, the Royal
Society and Leverhulme Trust, United Kingdom; DOE and NSF, United States of
America.

The crucial computing support from all WLCG partners is acknowledged
gratefully, in particular from CERN and the ATLAS Tier-1 facilities at
TRIUMF (Canada), NDGF (Denmark, Norway, Sweden), CC-IN2P3 (France),
KIT/GridKA (Germany), INFN-CNAF (Italy), NL-T1 (Netherlands), PIC (Spain),
ASGC (Taiwan), RAL (UK) and BNL (USA) and in the Tier-2 facilities
worldwide.


\bibliographystyle{atlasnote}
\bibliography{ditau}

\clearpage
\onecolumn
\begin{flushleft}
{\Large The ATLAS Collaboration}

\bigskip

G.~Aad$^{\rm 48}$,
B.~Abbott$^{\rm 110}$,
J.~Abdallah$^{\rm 11}$,
S.~Abdel~Khalek$^{\rm 114}$,
A.A.~Abdelalim$^{\rm 49}$,
A.~Abdesselam$^{\rm 117}$,
O.~Abdinov$^{\rm 10}$,
B.~Abi$^{\rm 111}$,
M.~Abolins$^{\rm 87}$,
O.S.~AbouZeid$^{\rm 157}$
% ...
\bigskip

% ...
$^{10}$ Institute of Physics, Azerbaijan Academy of Sciences, Baku, Azerbaijan\\
$^{11}$ Institut de F\'isica d'Altes Energies and Departament de F\'isica de la Universitat Aut\`onoma  de Barcelona and ICREA, Barcelona, Spain\\
% ...
$^{48}$ Fakult\"{a}t f\"{u}r Mathematik und Physik, Albert-Ludwigs-Universit\"{a}t, Freiburg i.Br., Germany\\
$^{49}$ Section de Physique, Universit\'e de Gen\`eve, Geneva, Switzerland\\
% ...
$^{87}$ Department of Physics and Astronomy, Michigan State University, East Lansing MI, United States of America\\
% ...
$^{110}$ Homer L. Dodge Department of Physics and Astronomy, University of Oklahoma, Norman OK, United States of America\\
$^{111}$ Department of Physics, Oklahoma State University, Stillwater OK, United States of America\\
% ...
$^{114}$ LAL, Univ. Paris-Sud and CNRS/IN2P3, Orsay, France\\
% ...
$^{117}$ Department of Physics, Oxford University, Oxford, United Kingdom\\
% ...
$^{157}$ Department of Physics, University of Toronto, Toronto ON, Canada\\
% ...
\end{flushleft}


\end{document}